\documentclass[preprint]{aastex}

\shortauthors{Ren et al.}
\usepackage{graphicx,multirow}
\usepackage{epstopdf}
\DeclareGraphicsExtensions{.eps}
\newcommand{\simlt}{\mathrel{\hbox{\rlap{\hbox{\lower4pt\hbox{$\sim$}}}\hbox{$<$}}}}
\newcommand{\simgt}{\mathrel{\hbox{\rlap{\hbox{\lower4pt\hbox{$\sim$}}}\hbox{$>$}}}}
\newcommand{\Ion}[2]{#1{\,\scriptsize #2}}
\usepackage[colorlinks, linkcolor=black,  anchorcolor=black, citecolor=black]{hyperref}
\shortauthors{Ren et al.}

\begin{document}

\title{WHITE-DWARF--MAIN-SEQUENCE BINARIES IDENTIFIED FROM THE LAMOST PILOT SURVEY}

\author{Juanjuan Ren\altaffilmark{1,2},
Ali Luo\altaffilmark{1,2},
Yinbi Li\altaffilmark{1},
Peng Wei\altaffilmark{1,2},
Jingkun Zhao\altaffilmark{1},
Yongheng Zhao\altaffilmark{1,2},
Yihan Song\altaffilmark{1},
Gang Zhao\altaffilmark{1,2}
}
\email{jjren@nao.cas.cn}
\email{lal@nao.cas.cn}

\altaffiltext{1}{Key Laboratory of Optical Astronomy, National Astronomical Observatories, Chinese Academy of Sciences, Beijing 100012, China}
\altaffiltext{2}{University of Chinese Academy of Sciences, Beijing, 100049, China}

\begin{abstract}
We present a set of white-dwarf--main-sequence (WDMS) binaries identified spectroscopically from the Large sky Area Multi-Object fiber Spectroscopic Telescope (LAMOST, also called the Guo Shou Jing Telescope) pilot survey. We develop a color selection criteria based on what is so far the largest and most complete Sloan Digital Sky Survey (SDSS) DR7 WDMS binary catalog and identify 28 WDMS binaries within the LAMOST pilot survey. The primaries in our binary sample are mostly DA white dwarfs except for one DB white dwarf. We derive the stellar atmospheric parameters, masses,  and radii for the two components of 10 of our binaries. We also provide cooling ages for the white dwarf primaries as well as the spectral types for the companion stars of these 10 WDMS binaries. These binaries tend to contain hot white dwarfs and early-type companions. Through cross-identification, we note that nine binaries in our sample have been published in the SDSS DR7 WDMS binary catalog. Nineteen spectroscopic WDMS binaries identified by the LAMOST pilot survey are new. Using the 3$\sigma$ radial velocity variation as a criterion, we find two post-common-envelope binary candidates from our WDMS binary sample.
\end{abstract}


\section{Introduction}\label{sec:intro}
White-dwarf--main-sequence (WDMS) binaries consist of a (blue) white dwarf primary and a (red) low-mass main-sequence (MS) companion formed from MS binaries where the primary has a mass $\lesssim 10\ M_{\Sun}$. Most of the WDMS binaries ($\sim 3/4$) are wide binaries \citep{silvestri2002, schreiber2010}, of which the initial MS binary separation is large enough that the two components will never interact and evolve like single stars \citep{de1992}. Consequently, the orbital period of these systems will increase because of the mass loss of the primary (the white dwarf precursor). The remaining $\sim 1/4$ of the WDMS binaries are close binaries. When the more massive star of a close binary leaves the MS and evolves into the giant branch or asymptotic giant branch phase, there is a dynamically unstable mass transfer to the MS companion and then the system goes through a common-envelope phase \citep{iben1993, zorotovic2010} during which it may continue evolving to an even shorter orbital period through angular momentum loss mechanisms (magnetic braking and gravitational radiation) or undergo a second common envelope. Binary population synthesis models \citep{willems2004} indicate the bimodal nature of the orbital period distribution of the entire population of WDMS binaries.

During the common-envelope phase of close binaries, the binary separation undergoes a rapid decrease and consequently orbital energy and angular momentum are extracted from the orbit, leading to the ejection of the envelope and exposing a post-common-envelope binary (PCEB). These PCEBs are important objects that may be progenitor candidates of cataclysmic variables \citep{warner1995} and perhaps Type Ia supernovae \citep{langer2000}. They can also improve the theory of close binary evolution \citep{schreiber2003}, especially the understanding of the physics of common-envelope evolution \citep{paczynski1976, zorotovic2010, rebassa2012b}. Among the variety of PCEBs (such as sdOB+MS binaries, WDMS binaries, and double degenerates), WDMS binaries are intrinsically the most common ones and the stellar components (white dwarfs and M dwarfs) of WDMS binaries are relatively simple. There are more and more WDMS binaries being found from the Sloan Digital Sky Survey \citep[SDSS;][]{york2000}, making them the most ideal population to help us understand common-envelope evolution. In addition, WDMS binaries, especially eclipsing WDMS binaries, also provide an interesting test of stellar evolution for both the white dwarf primary and the secondary low-mass MS in the binary environment \citep{nebot2009, parsons2010, parsons2011, parsons2012, parsons2013, pyrzas2009, pyrzas2012}.

Until now, a large number of WDMS binaries \citep{raymond2003, silvestri2007, heller2009, liu2012, rebassa2012a} have been efficiently identified from the SDSS. \citet{raymond2003} first attempted to study the WDMS pairs using SDSS data from the Early Data Release \citep{stoughton2002} and Data Release One \citep{abazajian2003}. They identified 109 WDMS pairs with $g <$ 20th magnitude. \citet{silvestri2006} presented 747 spectroscopically identified WDMS binary systems from the SDSS Fourth Data Release \citep{adelman2006}, and a further 1253 WDMS binary systems \citep{silvestri2007} from the SDSS Data Release Five \citep{adelman2007}. \citet{heller2009} identified 857 WDMS binaries from the SDSS Data Release Six \citep{adelman2008} through a photometric selection method. \citet{liu2012} identified 523 WDMS binaries from the SDSS Data Release Seven \citep[DR7;][]{abazajian2009} based on their optical and near-infrared color-selection criteria. \citet{rebassa2012a} provided a final catalog of 2248 WDMS binaries identified from the SDSS DR7, among which approximately 200 strong PCEB candidates have been found \citep{rebassa2012a}, and they further developed a publicly available interactive online database \footnote[1]{http://www.sdss-wdms.org/} for these spectroscopic SDSS WDMS binaries.

With these cumulative rich SDSS WDMS binary samples and the identified PCEBs, some important work has been done to test binary population models as well as study the stellar structures, magnetic activity, common-envelope evolution, and other properties. For example, \citet{nebot2011} pointed out that 21\%--24\% of SDSS WDMS binaries have undergone common-envelope evolution, which is in good agreement with the predictions of binary population models \citep{willems2004}. \citet{rebassa2011} found that the majority of low-mass ($M\mathrm{_{wd} \lesssim}$ 0.5 $M_{\Sun}$) white dwarfs are formed in close binaries by investigating the white dwarf mass distributions of PCEBs. \citet{zorotovic2012} studied the apparent period variations of eclipsing PCEBs and provided two possible  interpretations: second-generation planet formation or variations in the shape of a magnetically active secondary star. \citet{rebassa2013} investigated the magnetic-activity--rotation--age relations for M stars in close WDMS binaries and wide WDMS binaries using what is so far the largest and most homogeneous sample of SDSS WDMS binaries \citep{rebassa2012a}. They found that M dwarfs in wide WDMS binaries are younger and more active than field M dwarfs and the activity of M dwarfs in close binaries is independent of the spectral type.

As discussed above, several previous studies had developed color selection criteria based on the model colors of binary systems in order to select WDMS binaries from the SDSS \citep{szkody2002, raymond2003, smolcic2004, eisenstein2006, silvestri2006, liu2012}. For example,  \citet{raymond2003} developed an initial set of photometric selection criteria ($u - g < 0.45$, $g -r < 0.70$, $r - i > 0.30$, $i - z > 0.40$ to a limiting magnitude of $g <$ 20) to identify binary systems using SDSS photometry and yielded reliable results. Here we present a sample of WDMS binaries also selected with a series of photometric criteria from the spectra of Large sky Area Multi-Object fiber Spectroscopic Telescope \citep[LAMOST, also called the Guo Shou Jing Telescope, GSJT;][]{cui2012, zhao2012} pilot survey \citep{luo2012, yang2012, carlin2012, zhang2012, chen2012}. We discuss the sample selection and provide stellar parameters of the individual components of some binary systems (e.g., effective temperature, surface gravity, radius, mass and cooling age for the white dwarf; effective temperature, surface gravity, metallicity, spectral type, radius, and mass of the M dwarf).

The structure of this paper is as follows. In Section 2, we describe our sample selection criteria. In Section 3, we present the method of the spectral decomposition of some our WDMS binaries and derive the parameters for the two constituents of the binaries. Section 4 contains the analysis of the spectroscopic parameters of the binaries and the possible PCEB candidates in our sample. Finally, a brief conclusion is provided in Section 5.

\section{Identification of WDMS binaries in the LAMOST pilot survey}\label{sec:data}
\subsection{Introduction of the LAMOST Pilot Survey}
LAMOST (GSJT) is a quasi-meridian reflecting Schmidt telescope \citep{cui2012} located at the Xinglong Observing Station in the Hebei province of China. A brief description of the hardware and associated software of LAMOST can be found in \citet{cui2012}, which is a dedicated review and technical summary of the LAMOST project. The optical system of LAMOST has three major components: a Schmidt correcting mirror Ma (Mirror A), a primary mirror Mb (Mirror B), and a focal surface. The Ma (5.72 m $\times$ 4.40 m) is made up of 24 hexagonal plane sub-mirrors and the Mb (6.67 m $\times$ 6.05 m) has 37 hexagonal spherical sub-mirrors. Both are controlled by active optics. LAMOST has a field of view as large as 20 deg$^2$, a large effective aperture that varies from 3.6 to 4.9 m in diameter (depending on the direction it is pointing), and 4000 fibers installed on the circular focal plane with a diameter of 1.75 m. It has 16 spectrographs and 32 CCD cameras (each spectrograph equipped with two CCD cameras of blue and red channels), so there are 250 fiber spectra in each obtained CCD image. The up-to-date complete lists of LAMOST technical and scientific publications can be found at \url{http://www.lamost.org/public/publication}

\begin{figure}
\epsscale{0.9}
\plotone{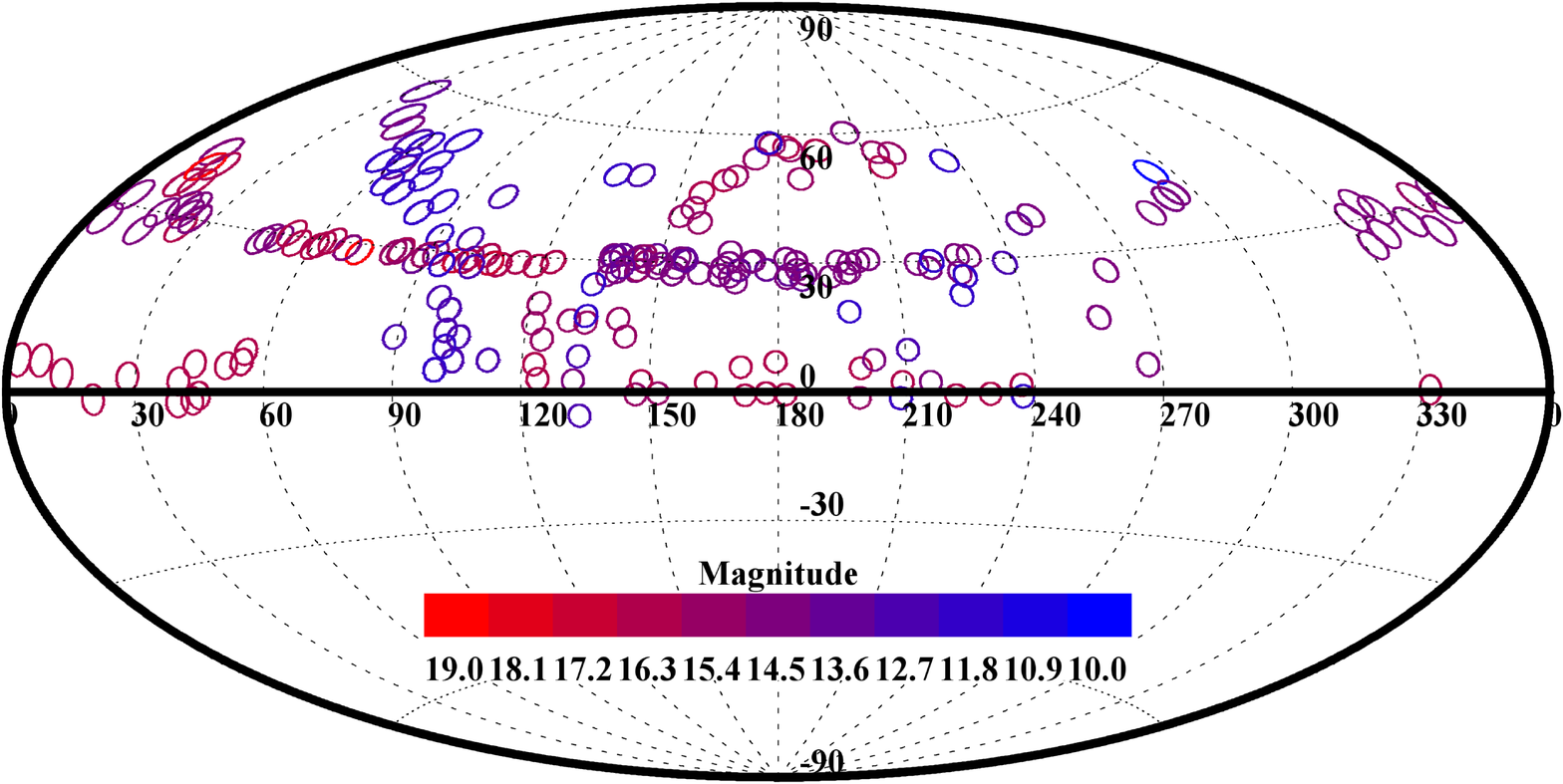}
\caption{Footprint of the LAMOST pilot survey in equatorial coordinates \citep{luo2012}. Each circle represents a plate of the pilot survey with radius 2$^{\circ}$.5}
\label{fig:footprint}
\end{figure}

The main aim of LAMOST is the extragalactic spectroscopic survey of galaxies (to study the large-scale structure of the universe) and the stellar spectroscopic survey of the Milky Way \citep[to study the structure and evolution of the Galaxy;][]{cui2012}. Based on these scientific goals, the LAMOST survey mainly contains two parts: the LAMOST ExtraGAlactic Survey (LEGAS) and the LAMOST Experiment for Galactic Understanding and Exploration (LEGUE) survey of the Milky Way (see \citet{deng2012} for the detailed science plan of LEGUE). It started a scientific spectroscopic survey of more than 10 million objects in 2012 September which will last for about five years. Before that, there was a two-year commissioning survey starting in 2009, and then a pilot spectroscopic survey performed using LAMOST from 2011 October to 2012 June. The LAMOST pilot survey is a test run of the telescope system to check instrumental performance and assess the feasibility of the science goals before the regular spectroscopic survey \citep{deng2012}. Based on the scientific goals proposed above, some scientists provided sources for LAMOST during the pilot survey. For example, \citet{chen2012} describe the LEGUE disk target selection for the pilot survey. There are about 380 plates \footnote[1]{http://data.lamost.org/pdr/plan} observed during the pilot survey, including the sources in the disk, spheroid, and anticenter of the Milky Way; M31 targets; galaxies; and quasars. Figure \ref{fig:footprint} shows the footprint of the LAMOST pilot survey \citep{luo2012}.

The LAMOST pilot survey includes nine full-Moon cycles during 2011 October and 2012 June (see \citet{yao2012} for details of the site conditions of LAMOST). Each moon cycle of the pilot survey was divided into three parts, dark nights (five days sequentially before and after the new moon), bright nights (five days sequentially before and after the full moon) and gray nights (the remaining nights in the cycle). In principle, dark nights and bright nights should be used for sources with magnitude 14.5 $< r <$ 19.5 and 11.5 $< r <$ 16.5 respectively, and the corresponding exposure times were 3 $\times$ 30 minutes and 3 $\times$ 10 minutes (depending on the distribution of the brightness of the sources), respectively. The raw CCD data are reduced and analyzed by the LAMOST data reduction system, which mainly includes the two-dimensional (2D) pipeline (for spectra extraction and calibration) and the one-dimensional (1D) pipeline \citep[to classify spectra and measure their redshifts;][]{luo2012}. What we want to mention is the flux calibration method of LAMOST, which is different from that of the SDSS. Because there is no network of photometric standard stars for LAMOST, \citet{song2012} proposed a relative flux calibration method. They select standard stars (like SDSS F8 subdwarf standards) for each spectrograph field. After observation, if the pre-selected standard stars do not have good spectral quality, they use the Lick spectral index grid method (see \citet{song2012} for details) to select new standard stars. Then the spectral response function for these standard stars is used to calibrate the raw spectra obtained from other fibers of the spectrograph \citep{song2012}.

About one million spectra were obtained during the LAMOST pilot survey (see \citet{luo2012} for the description of the data release of the LAMOST pilot survey). The LAMOST spectra have a low resolving power of $R \sim$ 2000 with wavelength ranging from 3800 $\mathrm{\AA}$ to 9000 $\mathrm{\AA}$. The LAMOST 1D FITS file is named in the form ``spec-MMMMM-YYYY$\_$spXX-FFF.fits", where ``MMMMM" represents the Modified Julian Date (MJD), ``YYYY" is the plan (or plate) identity number, ``XX" is the spectrograph identity number, and ``FFF" is the fiber identity number. After the reduction of the 1D pipeline, the spectra are classified (flag ``class") as either a galaxy, QSO, or star. For the spectra with low quality, we cannot guarantee that the classification given by the 1D pipeline is correct, so we give another flag ``final$\_$class'' based on the signal-to-noise ratio (S/N). For spectra with S/N $>$ 5, the spectral type of ``final$\_$class" is the classification result of the 1D pipeline, while for the spectra with low S/N, the ``final$\_$class" uses the visual inspection results. The flags ``subclass" and ``final$\_$subclass" are the spectral types of stars. The ``subclass" is the classification result of the 1D pipeline and the definition of ``final$\_$subclass" is the same as ``final$\_$class".

\subsection{WDMS Sample Selection}
Here we present a compilation of $\sim$ 28 WDMS binary systems from the LAMOST pilot survey. In order to search as many binaries as possible from the LAMOST pilot survey, we make use of the spectra with the ``final$\_$class" flag ``star" (644,540 stars) and the ``final$\_$subclass" belonging to spectral type ``M", ``K", ``WD", or ``Binary". After performing these spectral type cuts, we obtained an initial sample set containing 207,596 stars. In order to remove stars that are unlikely WDMS binary candidates from our 207,596 star sample, we plan to reduce the sample by means of the photometric selection method. The LAMOST survey is a multi-object spectroscopic survey that does not conduct an imaging survey like the SDSS. Because of the diversity of the target selection of the pilot survey, not all stars in the LAMOST pilot survey have SDSS magnitudes (by cross matching with the SDSS DR9 photometric catalog we find that about 60\% of the 644,540 stars have SDSS magnitudes), so we cannot use the SDSS $ugriz$ magnitude to do color selection and need to develop our own color selection criteria. We convolve the LAMOST spectra with the SDSS $ugriz$ filter response curves to obtain our own LAMOST magnitudes ($u^Lg^Lr^Li^Lz^L$) and colours ($u^L - g^L$, $g^L - r^L$, $r^L - i^L$, $i^L - z^L$).The LAMOST magnitudes we used in this paper are calculated by
\begin{equation}
m = -2.5 \log_{10} (F \otimes R),
\end{equation}
where, $m$ and $R$ represent the magnitude and response of the $ugriz$ filter and $F$ is the flux of the spectrum. For every observing night, we use the stars that have corresponding SDSS fiber magnitudes to fit a linear relationship between the SDSS fiber color and our convolved color to roughly calibrate our convolved color.

Because the SDSS WDMS binary catalog (2248 binaries) has a completeness of nearly $\gtrsim$ 98 $\%$ and represents what is so far the largest and most homogeneous sample set of WDMS binaries \citep{rebassa2012a}, we convolve these 2248 binary spectra with the SDSS $ugriz$ responsing curves (the same as the above procedure of calculating LAMOST convolved color) to help us give the color-cut criteria for the LAMOST sample. Considering that the spectra in the $u^L$ and $z^L$ bands have very small overlap with the photometric $u$ and $z$ band, we neglect the $u^L$ and $z^L$ band photometry and only use $g^L - r^L$ versus $r^L - i^L$ to constrain the color selection criteria. The following are color-cuts we adopted in this paper:
\begin{equation}
\begin{array}{lcllcl}
g - r & < & 0.45 + 1.59(r - i), & g -r & > & -0.61 + 0.01(r - i), \\
g - r & > & -1.92 + 1.84(r - i), & g - r & < & 2.02 - 0.38(r - i).
\end{array}
\end{equation}

\begin{figure}
\epsscale{1.1}
\plotone{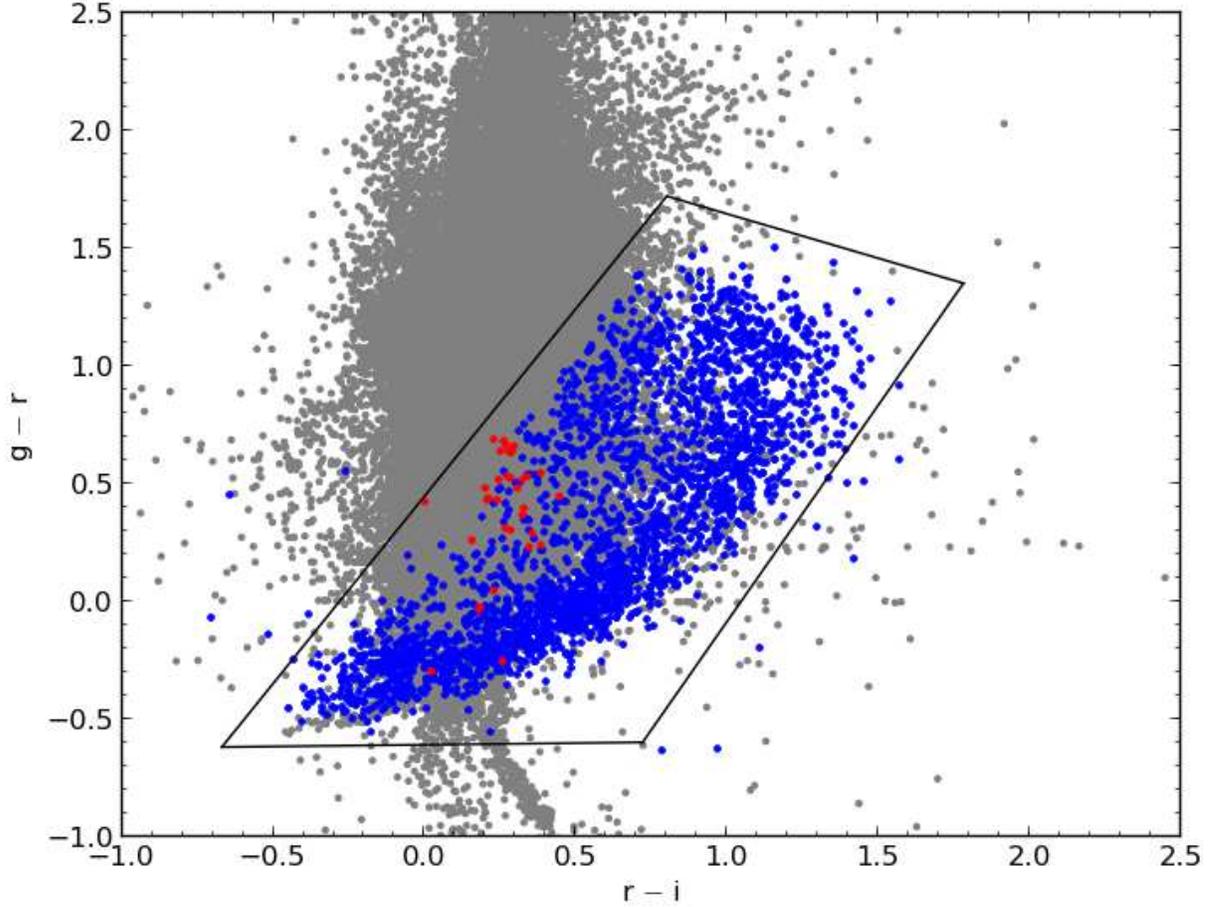}
\caption{Our WDMS binary colors in the $(g^{L} - r^{L})$ vs. $(r^{L} - i^{L})$ color--color diagrams. The area formed by the black lines shows the color-cuts used to select WDMS binaries within the LAMOST pilot survey. Colors for all stars in the LAMOST pilot survey and SDSS DR7 binaries given by \citet{rebassa2012a} are shown as gray and blue dots, respectively. Our resulting 33 WDMS spectra (corresponding to 28 WDMS binaries) are represented by red points.}
\label{fig:colorcut}
\end{figure}

After applying these color-cut criteria (the area formed by the black lines in Figure \ref{fig:colorcut}) to the 207,596 stars, our sample shrank to 90,079. We then searched for WDMS binary candidates among these 90,079 stars by eye check. The efficiency of the LAMOST blue spectrograph is much lower than the red one \citep{luo2012}, so for the spectra with low S/Ns in the blue band, the Balmer line series are easily superimposed by the noise. But the red band of the spectra have clear M dwarf line features for most of the samples. Based on these considerations, we select the WDMS binary candidates that clearly exhibit an M dwarf in the red band and a resolved white dwarf component with Balmer lines in the blue band of their spectra. In order to identify binaries easily, we decompose the spectra using principal component analysis (PCA) (see \citet{tu2010} for the details of the PCA decomposition). The WDMS binary spectrum is a mixture of a white dwarf spectrum and a M dwarf spectrum,
\begin{equation}
D \approx \sum _{i=1} ^{m} a_i w_i + \sum _{j=1} ^{n} a_{m+j} m_j,
\end{equation}
where, $D$ is the mixed spectrum, $w_i (i=1, 2, \dots, m)$ and $m_j (j=1, 2, \dots, n)$ represent the $m$ white dwarf eigen-spectra, and the $n$ M dwarf eigen-spectra we used to do spectral decomposition, respectively, and $a_i (i=1, 2, \dots, m+n)$ are the coefficients of eigen-spectra. At first, we construct four PCA eigen-spectra ($m=4$) from the white dwarf model spectral library and 12 PCA eigen-spectra ($n=12$) from the M dwarf model spectral library (the details of the white dwarf and M dwarf models can be found in Section 3). Then we determine all the coefficients of the eigen-spectra by the SVD (singular value decomposition) matrix decomposition and orthogonal transformation. After the coefficients of the eigen-spectra have been calculated, the spectrum can be decomposed into two components. By visual investigation of the decomposed spectra of the 90,079 stars decomposed by this PCA method, 33 WDMS binaries from the LAMOST pilot survey are found, of which 3 binaries have been observed more than once by LAMOST (J101616.82+310506.5 has been observed four times, J111035.16+280733.2 and J085900.86+493519.8 have each been observed twice). After eliminating the five duplicated spectra, our final sample includes 28 WDMS binaries (listed in Table 1) with only one DB white-dwarf--M-star binary (J224609.42$+$312912.2). Nine binaries of our WDMS sample have been identified by \citet{rebassa2012a} from the SDSS DR7.

Our 28 WDMS binaries show two components in their spectra. In order to ensure the reliability of our binary sample, additional investigations are carried out. For every binary pair in our sample, we check the spectra of its neighboring four fibers to see if there are fiber cross-contaminations brought about by the 2D data reduction pipeline \citep{luo2012}. For example, the neighboring spectra of the binary spectrum spec-55859-F5902$\_$sp08-169.fits (J221102.56$-$002433.5
) are spec-55859-F5902$\_$sp08-167.fits, spec-55859-F5902$\_$sp08-168.fits, spec-55859-F5902$\_$sp08-170.fits, and spec-55859-F5902$\_$sp08-171.fits. Their corresponding spectral types are SKY, A0, K7, and F5, respectively. The white dwarf component of this binary has very wide Balmer lines and may not be contaminated by an A0 type neighborhood. Meanwhile, the secondary component shows obvious molecule absorption band features of a M-type star, which also cannot be contaminated by the neighboring K7-type star. So this binary may not suffer from fiber contaminations. For other binaries, if their neighboring spectra have the same spectral types as their binary components, we will use the 2D image for further inspection. We do this check for each binary of our data and find that there are no fiber contaminations in our binary sample.

\begin{figure}
\epsscale{0.95}
\plotone{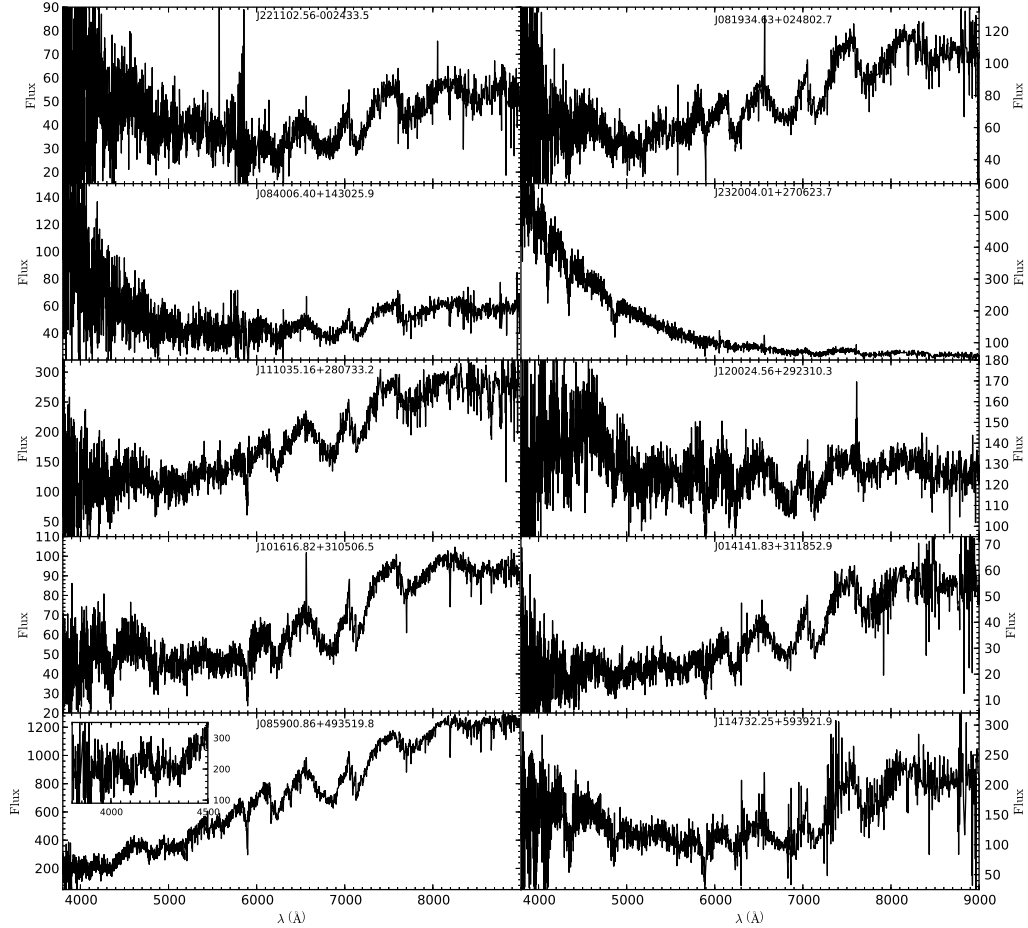}
\caption{Spectra of the 10 WDMS binaries for which we provided stellar parameters in this paper. For LAMOST J085900.86+493519.8, the blue component (white dwarf) is much weaker than the red component (M dwarf). To see the Balmer lines clearly, we insert a zoomed panel of the blue part of the spectrum (bottom left).}
\label{fig:spectra}
\end{figure}

\begin{figure}
\epsscale{1.1}
\plotone{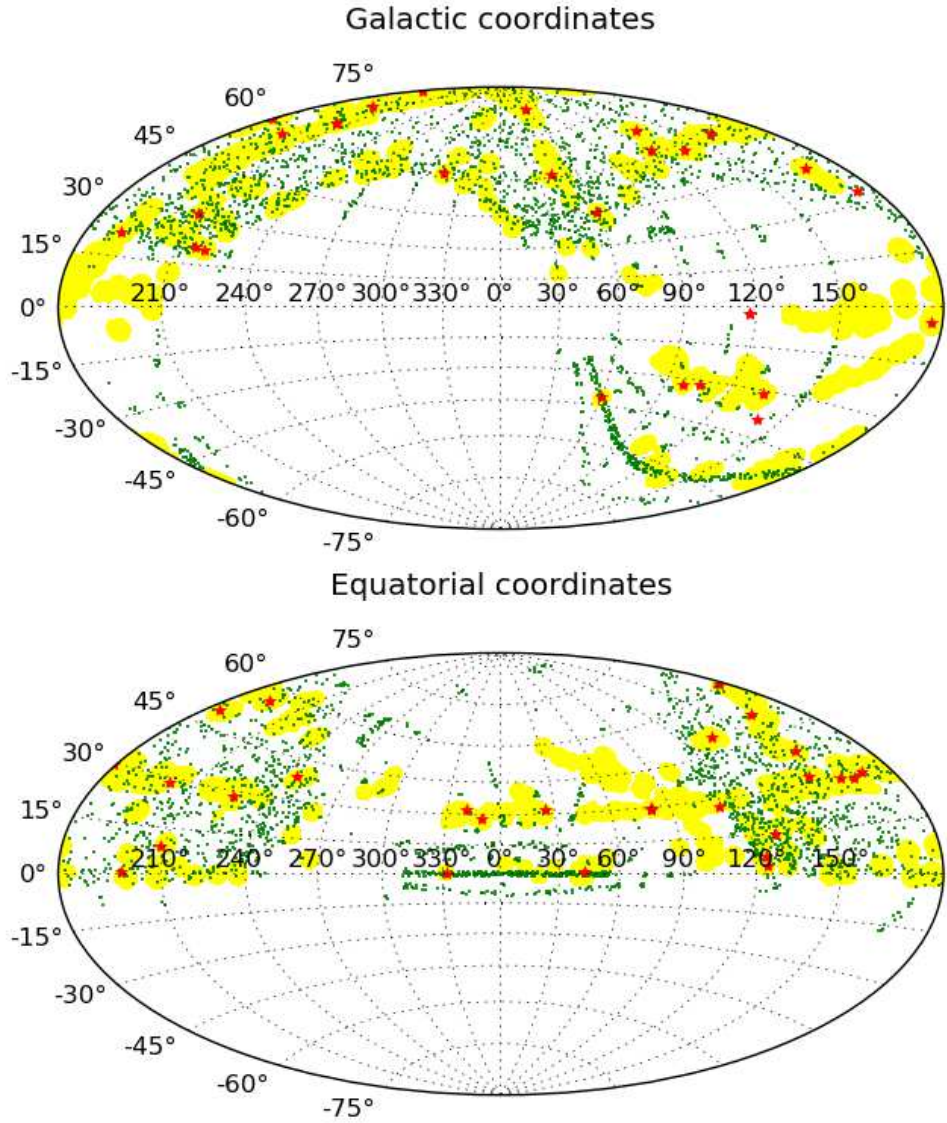}
\caption{Positions of our WDMS binary sample (red stars), SDSS DR7 binaries (green dots), and all stars of the LAMOST pilot survey (yellow dots) in the Galactic and equatorial coordinates.}
\label{fig:coords}
\end{figure}

Table 1 lists the colors of our convolved magnitudes and photometry for our sample cross matched with SDSS DR9 \citep{ahn2012}, 2MASS \citep{skrutskie2006}, and GALEX \citep{morrissey2007}. The Modified Julian Date (MJD), the plate identity number (PLT), the spectrograph identity number (SPID), and the fiber identity number (FIB) are also provided in Table 1. Figure \ref{fig:spectra} shows the 10 spectra of our binary sample on which we do spectral analysis in Section 3 (For the WDMS binaries that have multiple spectra, we only select one spectrum that has relatively good spectral quality to do the spectral analysis). Figure \ref{fig:coords} shows the positions of our  WDMS binary sample, the SDSS DR7 binaries \citep{rebassa2012a}, and all stars of the LAMOST pilot survey in the Galactic and equatorial coordinates.
\section{Stellar parameters}
To investigate the properties of the individual components of our WDMS binary systems, we adopt the template-matching method based on model atmosphere calculations for white dwarfs and M main-sequence stars to separate the two stellar spectra and derive parameters for each system. We only give stellar parameters for the 10 WDMS spectra with relatively high spectral quality (S/N $>$ 5). With the exception of one DB WDMS binary, the remaining 17 binaries not analyzed in this paper are either too noisy or insufficiently flux calibrated for a reasonable spectral analysis. After a good follow-up spectroscopic survey, we will provide stellar parameters for these binaries. The models we used for spectral decomposition and fitting and the parameter estimations for the two constituents are described in the following subsections.

\subsection{Models}
The theoretical model grids we used provide a five-dimensional parameter space $(\mathrm{T_{eff}^{WD}}$, $\mathrm{log (g_{WD})}$, $\mathrm{T_{eff}^M}$, $\mathrm{log(g_M)}$, $\mathrm{[Fe/H]_M})$ of white dwarfs and M main-sequence stars. Since most WDMS binaries contain a DA white dwarf, the white dwarf model spectra we used were calculated for pure hydrogen atmospheres (DA white dwarfs) based on the model atmosphere code described by \citet{koester2010}, which covers surface gravities of 7$\leq \mathrm{log}(g_\mathrm{WD}) \leq$ 9 with a step size of 0.25 dex and effective temperatures of 6000 K $\leq T_{\mathrm{eff}}^{\mathrm{WD}} \leq$ 90000 K. The step is 500 K, 1000 K, 2000 K, and 5000 K, respectively, when the effective temperatures are in the range of $[$6000 K, 15000 K$]$, $[$15000 K, 30000 K$]$, $[$30000 K, 50000 K$]$, and $[$50000 K, 90000 K$]$. The MARCS models \citep{gustafsson2008} are used as our MS model grids with a surface gravity range of 3.5 $\leq \mathrm{log}(g_\mathrm{M}) \leq$ 5.5 with step 0.5 dex, the effective temperature range 2500 K $\leq T_{\mathrm{eff}}^{\mathrm{M}} \leq$4000 K with step 100 K, and a metallicity of $\mathrm{[Fe/H]_M} \in \lbrace$ $-$2.0, $-$1.5, $-$1.0, $-$0.75, $-$0.5, $-$0.25, 0.0, 0.25, 0.5, 0.75, 1.0 $\rbrace$.

\subsection{Methods}
We use a template-matching method like \citet{heller2009} to decompose a WDMS binary spectrum into a white dwarf and a MS star simultaneously and derive independent parameters for each component. This method \citep{heller2009} is able to avoid a mutual dependence of the two scaling factors for WD and M stars since the system of equations can be solved uniquely and also avoids the effects of identifying a local $\chi ^2$ minimum. Our template matching method is not a bona fide weighted $\chi ^2$ minimization technique because we do not divide by $\sigma _i ^2$ (the observational error) in the $\chi ^2$ function of \citet[Equation(3)]{heller2009}. By comparing the fitting method divided by $\sigma _i ^2$ with the method that is not, we note that the template-fitting method that does not divide by $\sigma _i ^2$ gives much better results. The main reason for this is that the observational error given by the LAMOST ``.fits" file is the inverse-variance of the flux and may not exactly represent the uncertainty of the flux, so it will distinctly affect the $\chi ^2$ fitting results.

\begin{figure}
\epsscale{1.1}
\plotone{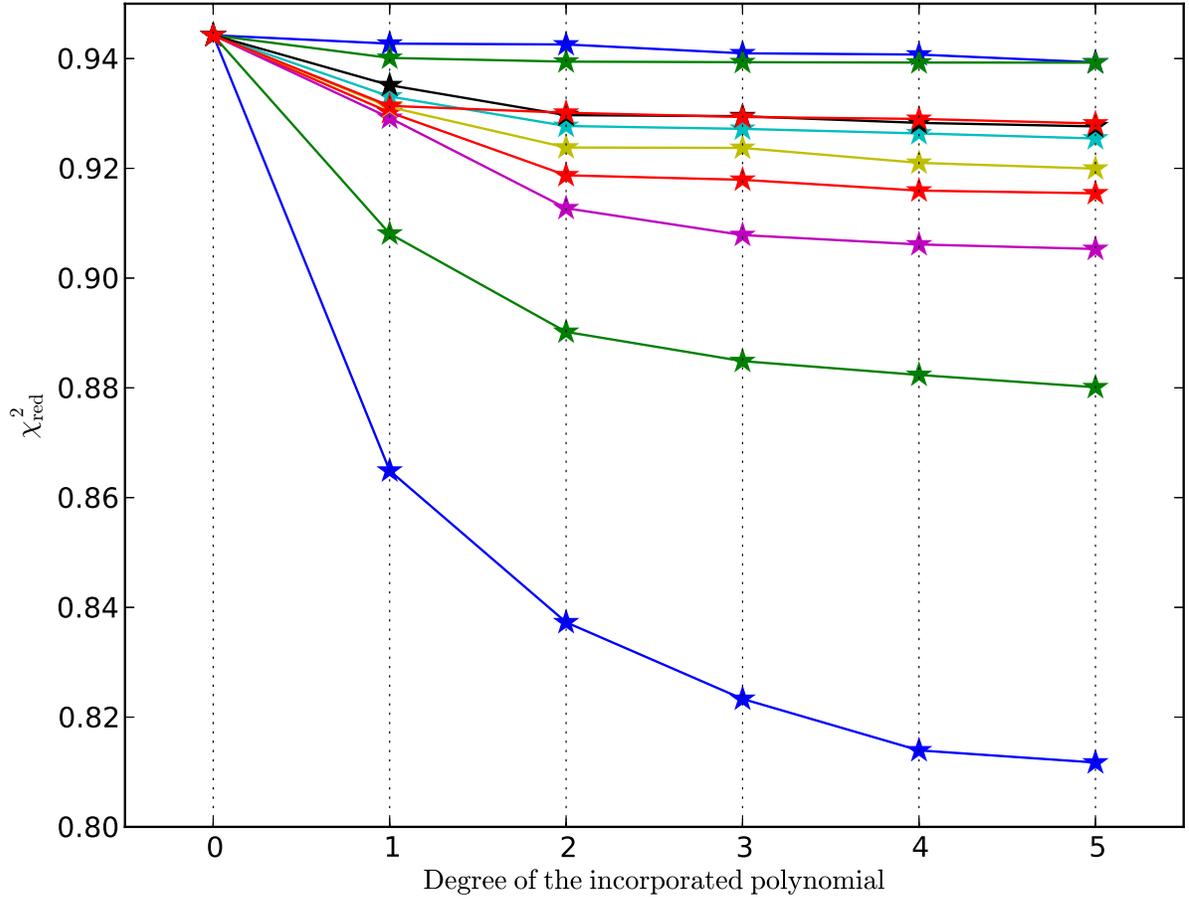}
\caption{ Reduced $\chi^{2}$ distribution of 10 WDMS binaries for which we do spectral analysis for the fitting routine with polynomials of different orders. The abscissa represents the degree of the polynomial, and the ordinate shows the reduced $\chi^2$ ($\chi^2 _{\mathrm{red}}$). For comparison, we shift all 10 lines to the same initial point (the $\chi^2 _{\mathrm{red}}$ for the fitting routine without the polynomial). }
\label{fig:poly}
\end{figure}

\begin{figure}
\epsscale{0.95}
\plotone{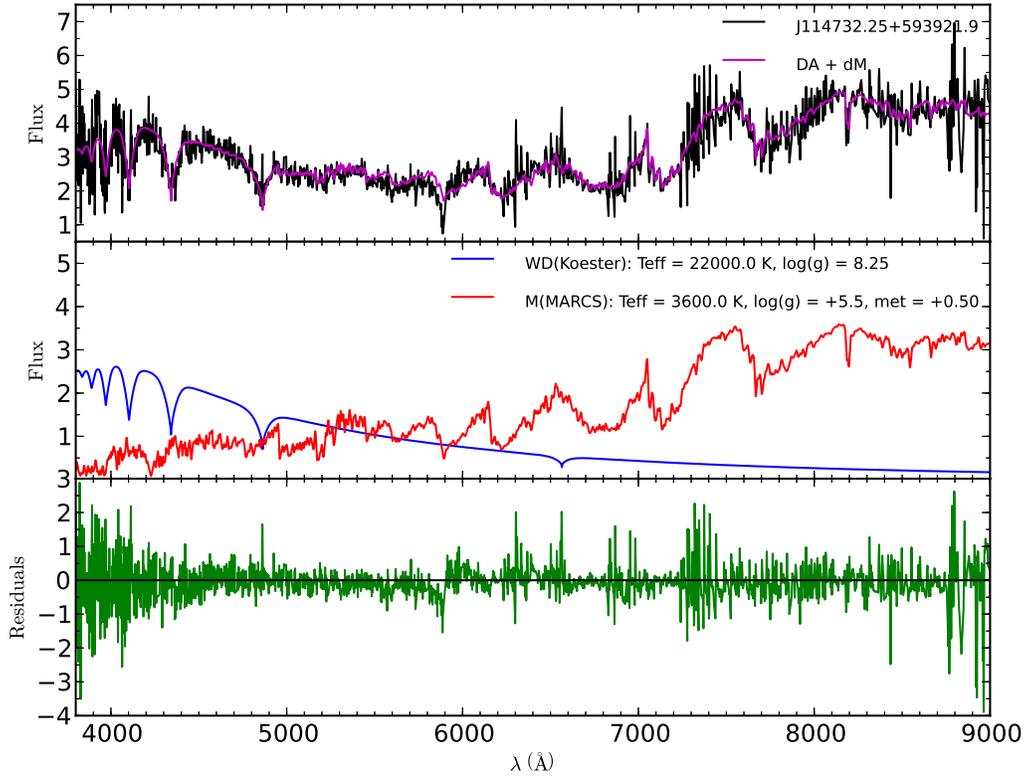}
\caption{Top: spectrum of the WDMS binary system LAMOST J114732.25+593921.9 and the fitting models obtained by our $\chi ^2$ minimization technique. Center: decomposed two components of the observed spectrum. Bottom: residuals between the LAMOST J114732.25+593921.9 spectrum and our fitting results.}
\label{fig:example}
\end{figure}

Additionally, the flux calibration of the LAMOST spectra is relative (mentioned in Section 2, or see details in \cite{song2012}), which means the reddening of the standard stars may affect the flux calibration, especially when the standard stars have very different spacial positions (or reddening) from other stars in the same spectrograph. This may lead to some difference in the continuum between the observational spectra and model spectra. Considering the complexity of reddening, to simplify we empirically incorporate quintic polynomials to Heller's template-matching method in order to overcome the possible failure of the fit. Therefore, the definition of $\chi^2$  changes to 
\begin{equation}
\chi ^2 = \sum _{i} ^{n} {\left(F_i - \left(\sum\limits_{j=0}^{5} P_j x_i ^j \right)w_i - \left(\sum\limits_{j=0}^{5} Q_j x_i ^j \right) m_i \right)^2},
\end{equation}
where $F_i$, $w_i$, and $m_i$ are the observed flux, white dwarf model flux, and M dwarf model flux for each data point in a binary spectrum with a total number of $n$ observed data points. $x_i$ is the wavelength of the spectrum. The quintic polynomials we incorporated are $\sum\limits_{j=0}^{5} P_j x_i ^j$ and $\sum\limits_{j=0} ^{5} Q_j x _i ^j$. Using this revised template-fitting method, we estimate the stellar parameters of the white dwarf component $\lbrace T\mathrm{_{eff}^{WD}}$, $\mathrm{log (}g\mathrm{_{WD})}\rbrace$ and its M-type MS companion $\lbrace T\mathrm{_{eff}^M}$, $\mathrm{log(}g\mathrm{_M)}$, $\mathrm{[Fe/H]_M}\rbrace$ for our binaries. To verify that the quintic polynomials are necessary, we incorporate polynomials of different orders (from zero to five) into our fitting routine. For comparison, we calculate the reduced $\chi ^2$ ($\chi ^2 _{\mathrm{red}}$) like \citet{heller2009}. Figure \ref{fig:poly} shows the $\chi ^2 _{red}$ distribution for the fitting routine with different order polynomials.  It seems that the $\chi^2 _{\mathrm{red}}$ converges when we incorporate five-order polynomials into the fitting routine. An example of a typical WDMS spectrum in our sample and its spectral decomposition are shown in Figure \ref{fig:example}.

As \citet{heller2009} mentioned, the quality of this spectra-fitting technique is poor in a mathematical context and the standard deviations of the measured parameters are quite weak in terms of physical significance. That means that the unknown systematic errors (due to the incomplete molecular data of the M star model, flux calibration errors, and possible interstellar reddening) of the stellar parameters are larger than the mathematical ones. For a conservative estimate of errors in the measured parameters, we refer to \citet{hugelmeyer2006} and assume an uncertainty of half the model step width for the WDs and MS stars. We assume an uncertainty of 2000 K for WDs with effective temperatures of less than 50,000 K and 100 K for the MS stars. While for $T\mathrm{_{eff} ^{WD}} > 50,000$ K, the uncertainty is given by half of the model step width of 5000 K. Limited by the low resolution of the LAMOST spectra and the step size of our model grid, the accuracy of the surface gravities is given as $\sigma_{\mathrm{log}(g_{\mathrm{WD}})} \approx 0.25$ dex and $\sigma_{\mathrm{log}(g_{\mathrm{M}})} \approx 0.5$ dex. Like \citet{heller2009}, we also assume an uncertainty of 0.3 dex for metallicity below $-$1.0 dex. For metallicity over $-$1.0 dex, the corresponding uncertainty is given by half of the model step width of 0.25 dex.

Because the flux-calibration of LAMOST spectra are relative, it is impossible for us to derive the distances of the two components of our WDMS binaries from the best-fitting flux scaling factors that scale the model flux to the observed flux. Once $T\mathrm{_{eff} ^{WD}}$ and $\mathrm{log}(g_{\mathrm{WD}})$ are determined, we estimate the cooling ages, masses and radii for the white dwarfs by interpolating the detailed evolutionary cooling sequences \citep{wood1995, fontaine2001}. We use the carbon-core cooling models \citep{wood1995} with thick hydrogen layers of $q_{H} = M_{H}/M_{*} = 10^{-4}$ for pure hydrogen model atmospheres with effective temperature above 30,000 K. For $T_{eff}$ below 30,000 K, we use cooling models similar to those described in \citet{fontaine2001} but with carbon-oxygen cores and $q_{H} = 10^{-4}$. Additionally, we derive the masses and radii for M-type companions using the empirical effective temperature -- spectral type ($T_{\mathrm{eff}}$ -- Sp), spectral type -- mass (Sp -- M) and spectral type --radius (Sp -- $R$) relations presented in \citet{rebassa2007}. These spectroscopic parameters are listed in Table 2.

\section{Results and Discussion}\label{sec:results}
\subsection{Analysis of Stellar Parameters}
Our WDMS binary sample is not complete and we only provide stellar parameters for 10 binaries, so the statistical distribution analysis of parameters is not given. The white dwarf effective temperatures are between 21,000 K and 48,000 K and have a mean value 29,900 K. This may suggest that this WDMS binary sample tends to have hot white dwarfs. The mean value of white dwarf surface gravities is around 8.0, which is consistent with the peak value of surface gravity distribution of \citet{rebassa2012a}. \citet{yi2013} give an M dwarf catalog from the LAMOST pilot survey, for which the spectral type has a peak around M1 $\sim$ M2. The spectral types of the secondary stars of our WDMS binaries also cluster together around M1.5. All of these imply that our sample favours binaries with hot white dwarfs and early-type companion stars. The white dwarf masses of the 10 binaries for which we provided stellar parameters tend to be higher than the typical peak of WD mass $\sim$ 0.6 $M_\Sun$ \citep{tremblay2011}. That may be because of the selection effect and the small sample size. Our binary sample is still not large enough for parameter distribution analysis and is waiting to be enlarged by the ongoing LAMOST formal survey.

\subsection{Radial Velocities}
\begin{figure}
\epsscale{1.1}
\plotone{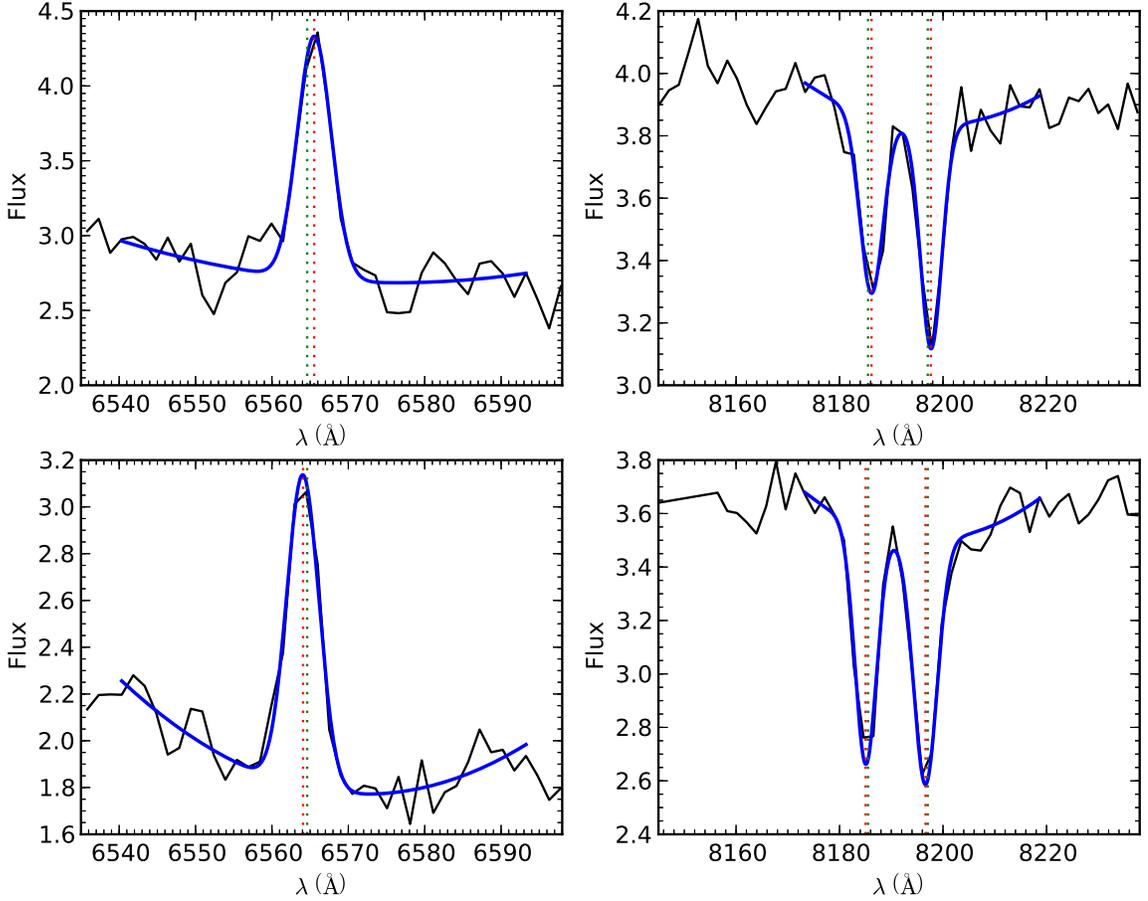}
\caption{Fitting results of the H$\alpha$ $\lambda$ 6564.61 emission line (left panel) and \Ion{Na}{I} $\lambda \lambda$ 8183.27, 8194.81 absorption doublet (right panel) with the Gaussian function of two LAMOST binaries J081934.63$+$024802.7 and J162617.40$+$384029.9. The H$\alpha$ line is fitted with a Gaussian function plus a parabola, and the \Ion{Na}{I} doublet is fitted with a double Gaussian function plus a parabola. The black line is the observational spectrum and blue line is the fitted Gaussian profile. The vertical green and red dotted lines indicate the center of H$\alpha$ emission line and \Ion{Na}{I} doublet in the rest frame and our fitting results, respectively.}
\label{fig:rv}
\end{figure}

For the binary spectra that exhibit the resolved spectral H$\alpha$ $\lambda$ 6564.61 emission line and \Ion{Na}{I} doublet $\lambda\lambda$ 8183.27, and 8194.81 lines, we try to derive radial velocities by fitting the H$\alpha$ emission line with a Gaussian line profile plus a second order polynomial as well as fitting the \Ion{Na}{I} doublet with a double Gaussian profile of fixed separation and a second order polynomial \citep{rebassa2007}. The total error of the radial velocities is computed by quadratically adding the uncertainty of the LAMOST wavelength calibration \citep[10 km s$^{-1}$; see][]{luo2012} and the error in the position of the H$\alpha$/\Ion{Na}{I} lines determined from the Gaussian fits. The LAMOST spectra are generally combined from three exposures (which we call `subspectra' following \citet{rebassa2007}), with each exposure lasting for either 30 minutes or 15 minutes (see Section 2). We then measure radial velocities for these LAMOST subspectra and the combined spectra. Last, we derive radial velocities for 17 binaries (see Table 3), of which J085900.86+493519.8 and J101616.82+310506.5 have been observed two and four times, respectively, by LAMOST. Figure \ref{fig:rv} shows the fitting results of the H$\alpha$ $\lambda$ 6564.61 emission line and the \Ion{Na}{I} doublet $\lambda\lambda$ 8183.27, and 8194.81 lines of two LAMOST spectra.

\subsection{PCEB Candidates}
From Table 3, 10 binaries have been measured with multiple radial velocities (including the radial velocities measured from LAMOST subspectra and provided by the SDSS WDMS binary catalog \citep{rebassa2012a}). Following \citet{rebassa2010} and \citet{nebot2011}, we consider those systems showing radial velocity variations with 3$\sigma$ significance to be strong PCEB candidates. We used a $\chi^2$ test with respect to the mean radial velocity for the detection of radial velocity variations. If the probability $Q$ that the $\chi^2$ test returns is below 0.0027 (meaning the probability $P(\chi^2)$ of a system showing large radial velocity variation is above 0.9973 where $P(\chi^2)=1-Q$), we can say that we detect 3$\sigma$ radial velocity variation and the corresponding WDMS binary can be considered a strong PCEB candidate. Here we find two PCEB candidates, LAMOST J105421.88$+$512254.1 and LAMOST J122037.01$+$492334.0, using the \Ion{Na}{I} doublet radial velocities. LAMOST J105421.88$+$512254.1 shows 3$\sigma$ radial velocity variation using either \Ion{Na}{I} doublet or H$\alpha$ emission, while for LAMOST J122037.01$+$492334.0, we detect 3$\sigma$ radial velocity variation using the \Ion{Na}{I} doublet. By cross-referencing  195 PCEB candidates provided by \citet{rebassa2012a}, we find that LAMOST J105421.88$+$512254.1 has already been identified as a PCEB candidate. Limited by the sample size, we only find one new PCEB candidate: LAMOST J122037.01$+$492334.0.

We estimate the upper limits to orbital periods for these two PCEB candidates in the same way as described in \citet{rebassa2007}. Because we do not provide stellar parameters for these two binaries, the white dwarf masses and the secondary star masses are taken from \citet{rebassa2012a}. The radial velocity amplitudes of the secondary stars are obtained by using the \Ion{Na}{I} doublet radial velocities (Table 3). Table 4 presents the probability $P(\chi^2)$ of measuring large radial velocity variations for these two binaries and the calculated upper limits to their orbital periods. Both PCEB candidates need intense follow-up spectroscopic observations in order to obtain orbital periods.

\section{Conclusions}\label{sec:conclusion}
We have presented a catalog of 28 WDMS binaries from the spectroscopic LAMOST pilot survey in this paper. Using the colors of the 2248 binaries from the SDSS WDMS binary catalog, we develop our own color selection criteria for the selection of WDMS binaries from the LAMOST pilot survey based on the LAMOST magnitudes obtained by convolving the LAMOST spectra with the SDSS $ugriz$ filter response curve. This method is efficient for searching for binaries in the spectroscopic survey without having our own optical or infrared photometric data equipment like the LAMOST survey. Using the color selection criteria, we identify 28 WDMS binaries from the LAMOST pilot survey. Nine of these binaries have been published in previous works and 19 of these WDMS binaries are new. For 10 of our binaries, we have used a $\chi^2$  minimization technique to decompose the binary spectra and determine the effective temperatures, surface gravities of the white dwarfs, as well as the effective temperatures, surface gravities, and metallicities for the M-type companions. The cooling ages, masses and radii of white dwarfs are provided by interpolating the white dwarf cooling sequences for the derived effective temperatures and surface gravities. We also derive the spectral types, masses, and radii of the M stars by using empirical spectral type -- effective temperature, spectral type -- mass, and spectral type -- radius relations. In addition, the radial velocities are measured for most of our sample. We also  discuss the possible PCEB candidates among the binary systems with multiple spectra in our sample, finally giving two possible PCEB candidates, one of which has already been identified as a PCEB candidate by the SDSS.

The WDMS binary catalog is the first provided by the LAMOST survey, and demonstrates the capability of LAMOST to search for WDMS binaries. With the ongoing formal LAMOST survey, we hope to find many more WDMS binaries and present a more complete catalog of WDMS binaries, which will increase the number of known WDMS binary systems. Enlarging the WDMS binary sample will lead to a deeper understanding of close compact binary star evolution and other follow-up studies.

\acknowledgments
We thank the anonymous referee for very useful comments and suggestions that greatly improved this paper. We thank Alberto Rebassa-Mansergas for the discussion about the detection of PCEBs. We also thank Detlev Koester for kindly providing the white dwarf atmospheric models and Bengt Edvardsson for useful discussion of the MARCS models. We acknowledge the Web site \url{http://www.astro.umontreal.ca/~bergeron/CoolingModels} for providing the white dwarf cooling sequences. We thank George Comte for the discussion of M dwarf models and Xiaoyan Chen, Yue Wu, and Fang Zuo for discussion. This study is supported by the National Natural Science Foundation of China under grant Nos. 10973021, 11078019, and 11233004. The Guo Shou Jing Telescope (the Large Sky Area Multi-Object Fiber Spectroscopic Telescope, LAMOST) is a National Major Scientific Project built by the Chinese Academy of Sciences. Funding for the project has been provided by the National Development and Reform Commission. LAMOST is operated and managed by the National Astronomical Observatories, Chinese Academy of Sciences. The LAMOST Pilot Survey Web site is \url{http://data.lamost.org/pdr}.

\clearpage
\onecolumn

\begin{deluxetable}{lccrrrrrrccccccccccc}
\rotate
\tablecolumns{18}
\tabletypesize{\scriptsize}
\setlength{\tabcolsep}{0.02in}
\tablewidth{-1pt}
\tablecaption{Photometric Properties of LAMOST WDMS Binaries}
\tablehead{
\colhead{Designation} & \colhead{R.A.} & \colhead{Decl.} &
\colhead{MJD} & \colhead{PLT} & \colhead{SPID} & \colhead{FIB} &
\colhead{$(g^L - r^L)$} & \colhead{$(r^L - i^L)$} &
\colhead{{\it u}} & \colhead{{\it g}} & \colhead{{\it r}} & \colhead{{\it i}} & \colhead{{\it z}} &
\colhead{{\it J}} & \colhead{{\it H}} & \colhead{{\it K}} &
\colhead{{fuv}} & \colhead{{nuv}} &
\colhead{Flag} \\
&(deg)&(deg)&&&&&&(mag)&(mag)&(mag)&(mag)&(mag)&(mag)&(mag)&(mag)&(mag)&(mag)&}
\startdata
J014141.83$+$311852.9 & 25.42433 & 31.3147 & 55918 & M31$\_$025N30$\_$M1 & sp15 & 033
 & 0.660554 & 0.270636 & 19.393 & 18.429 & 17.53 & 16.446 & 15.826 & 14.522 & 13.816 & 13.619 & 19.521 & 19.581 & \\
J025306.36$+$001329.6 & 43.2765 & 0.22491 & 55859 & F5907 & sp14 & 137
 & 0.509917 & 0.248582 & 19.359 & 18.916 & 19.308 & 19.818 & 20.025 &	 14.173 & 13.546 & 13.272 & 19.601 & 19.429 & e,r \\
J052529.17$+$283705.4 & 81.37156 & 28.61819 & 55859 & F5909 & sp04 & 114
 & 0.360057 & 0.324585 & ... & ... & ... & ... & ... & 15.354 & 14.596 & 14.368 & ... & ... & \\
J052531.26$+$283807.6 & 81.38025 & 28.63545 & 55951 & GAC$\_$080N28$\_$M1 & sp04 & 114
 & 0.280857 & 0.361304 & ... & ... & ... & ... & ... & 13.401 & 12.77 & 12.553 & ... & ... & \\
J052531.33$+$284549.4 & 81.38058 & 28.76375 & 55859 & F5909 & sp04 & 120
 & 0.387878 & 0.332709 & ... & ... & ... & ... & ... & 15.583 & 15.027 & 14.71 & ... & ... & \\
J073128.30$+$264353.8 & 112.86793 & 26.73163 & 55921 & GAC$\_$113N28$\_$M1 & sp01 & 044
 & 0.510806 & 0.327412 & 18.46 & 17.772 & 17.248 & 16.204 & 15.498 & 14.049 & 13.459 & 13.283 & 19.6 & 18.689 & \\
J081934.63$+$024802.7 & 124.8943 & 2.80077 & 55921 & F5592103 & sp08 & 044
 & 0.631215 & 0.293536 & 17.721 & 17.255 & 16.561 & 15.686 & 15.074 & 13.708 & 13.076 & 12.897 & ... & ... & \\
J081959.21$+$060424.2 & 124.996713 & 6.073391 & 55892 & F9205 & sp06 & 053
 & 0.444555 & 0.446157 & 19.665 & 19.433 & 19.456 & 18.811 & 18.154 & 16.58 & 16.357 & 15.847 & 19.188 & 19.421 & r \\
J084006.40$+$143025.9 & 130.02669 & 14.50722 & 55977 & F5597707 & sp01 & 145
 & 0.304351 &	 0.269564 & 18.048 & 18.109 & 17.996 & 17.318 & 16.819 & 15.447 & 14.934 & 14.594 & ... & ... & \\
J085900.86$+$493519.8 & 134.75359 & 49.58885 & 56021 & VB3$\_$136N49$\_$V1 & sp03 & 087
 & $-$0.034498 & 0.182911 & 16.466 & 15.049 & 13.9 & 13.297 & 12.539 & 11.251 & 10.577 & 10.364 & 16.736 & 16.678 & \\
J101616.82$+$310506.5 & 154.07012 & 31.08516 & 55907 & B90705 & sp03 & 065
 & 0.634328 & 0.283063 & 17.38 & 16.751 & 16.167 & 15.317 & 14.787 & 13.47	& 12.885 & 12.698 & 17.296 & 17.241 & \\
J103514.80$+$390717.7 & 158.8117 & 39.1216 & 55930 & F5593001 & sp04 & 124
 & 0.425026 & 0.208033 & 18.423 & 18.178 & 17.919 & 17.301 & 16.859 & 15.565 & 15.034 & 14.849 & 17.768 & 18.07 & s,r \\
J105405.25$+$283916.7 & 163.52188 & 28.65466 & 55931 & B5593104 & sp16 & 214
 & 0.519861 & 0.284815 & 17.824 & 17.146 & 16.645 & 15.477 & 14.657 &   13.202 & 12.569 & 12.339 & 18.024 & 17.794 & r \\
J105421.88$+$512254.1 & 163.59118 & 51.38171 & 55915 & F5591506 & sp11 & 205
 & 0.526326 & 0.275433 & 17.258 & 16.778 & 16.134 & 15.271 & 14.729 & 13.455 & 12.813 & 12.575 & 16.46 & 16.817 & s,r \\
J111035.16$+$280733.2 & 167.64653 & 28.1259 & 55931 & B5593104 & sp13 & 180
 & 0.687018 & 0.232178 & 17.786 & 17.161 & 16.427 & 15.554 & 15.024 & 13.729 & 13.099 & 12.911 & 17.371 & 17.506 & \\
J112512.36$+$290545.4 & 171.30153 & 29.09597 & 56018 & B5601803 & sp02 & 195
 & 0.228522 & 0.345777 & 18.817 & 18.215 & 17.521 & 16.481 & 15.856 & 14.531 & 13.907 & 13.635 & 18.474 & 18.611 & r \\
J114732.25$+$593921.9 & 176.8844 & 59.65611 & 55931 & F5593101 & sp11 & 001
 & 0.424084 & 0.243105 & 17.127 & 16.796 & 16.474 & 15.514 & 14.882 & 13.485 & 12.832 & 12.635 & 16.682 & 16.783 &	\\
J120024.56$+$292310.3 & 180.10235 & 29.3862 & 55961 & B5596106 & sp02 & 085
 & 0.432907 & 0.216112 & 20.275 & 17.715 & 16.386 & 15.665 & 15.216 & 14.068 & 13.44 & 13.273 & ... & ... & \\
J122037.01$+$492334.0 & 185.1542208 & 49.3927944 & 55923 & F5592306 & sp03 & 077
 & 0.63171 & 0.255231 & 18.719 & 18.14 & 17.417 & 16.809 & 16.405 & 15.172 & 14.505 & 14.396 & ... & ... & s,r \\
J125555.30$+$560033.0 & 193.98043 & 56.00917 & 55952 & F5595206 & sp04 & 123
 & 0.623711 & 0.288381 & 22.695 & 20.013 & 18.701 & 17.558 & 16.898 & 15.682 & 15.063 & 15.027 & ... & ... & \\
J131208.11$+$002058.0 & 198.0338083 & 0.3494583 & 55932 & F5593201 & sp16 & 071
 & 0.0403807 & 0.232714 & 18.79 & 18.472 & 18.35 & 17.564 & 16.927 & 15.625 & 15.029 & 14.787 & 18.247 & 18.53 & r,c \\
J132417.76$+$280755.8 & 201.07402 & 28.13217 & 55975 & B5597505 & sp01 & 195
 & 0.477785 & 0.203373 & 17.477 & 16.903 & 16.524 & 15.593 & 14.979 & 13.57 & 12.941 & 12.658 & 17.325 & 17.378 & \\
J135635.32$+$084841.7 & 209.1472 & 8.81159 & 56062 & VB3$\_$210N09$\_$V2 & sp10 & 161
 & 0.544101 & 0.38414 & 17.524 & 16.798 & 16.057 & 15.021 & 14.403 & 13.042 & 12.416 & 12.169 & 17.387 & 17.417 & \\
J150626.53$+$275925.2 & 226.61058 & 27.99036 & 56063 & B5606303 & sp09 & 122
 & $-$0.260791 & 0.260247 & 18.258 & 18.177 & 18.053 & 17.5 & 17.043 & 15.852 & 15.253 & 14.698 & 17.347 & 17.602 & r \\
J162617.40$+$384029.9 & 246.57251 & 38.67499 & 55999 & B5599908 & sp15 & 040
 & 0.253555 & 0.159056 & 17.361 & 16.938 & 16.697 & 15.7 & 15.111 & 12.513 & 11.947 & 11.663 & ... & ... & \\
J221102.56$-$002433.5 & 332.76069 & $-$0.40933 & 55859 & F5902 & sp08 & 169
 & 0.477924 & 0.309289 & 18.634 & 18.307 & 17.941 & 16.993 & 16.415 & 15.154 & 14.501 & 14.352 & 18.213 & 18.466 & \\
J224609.42$+$312912.2 & 341.53925 & 31.48674 & 55863 & B6302 & sp09 & 187
 & 0.235289 & 0.387328 & 16.952 & 15.671 & 15.19 & 15.018 & 14.936 & 14.11 & 13.728 & 13.584 & ... & 20.489 & \\
J232004.01$+$270623.7 & 350.01673 & 27.10659 & 55878 & B87802$\_$1 & sp15 & 083
 & $-$0.305056 & 0.0273775 & 15.852 & 16.072 & 16.378 & 16.168 & 15.811 & 14.609 & 14.074 & 13.783 & 14.747 & 15.203 &
\enddata
\tablecomments{
 Coordinates and SDSS DR9 ($u$, $g$, $r$, $i$, $z$), 2MASS ($J$, $H$, $K$), and GALEX (fuv, nuv) magnitudes for our WDMS binary sample are included. Columns (4)--(6) list the LAMOST color $(g^L - r^L)$ and $(r^L - i^L)$ . We use `...' to indicate that no magnitude is available. J224609.42$+$312912.2 is the only DB white dwarf--main-sequence binary in our sample. Flags e, s, c, and r represent those binaries which have been studied by \citet{eisenstein2006}, \citet{silvestri2006}, \citet{schreiber2010}, \citet{rebassa2010}, and \citet{rebassa2012a}.}
\end{deluxetable}

\begin{deluxetable}{lccccccccccc}
\rotate
\tablecolumns{17}
\tabletypesize{\scriptsize}
\setlength{\tabcolsep}{0.02in}
\tablewidth{-1pt}
\tablecaption{Spectroscopic Properties of LAMOST WDMS Binaries.}
\tablehead{
\colhead{Designation} &
\colhead{$T\mathrm{_{eff} ^{WD}}$} & \colhead{$\mathrm{log}(g\mathrm{_{WD}})$} &
\colhead{$T\mathrm{_{eff} ^{M}}$} & \colhead{$\mathrm{log}(g\mathrm{_{M}})$} &
\colhead{[Fe/H]$\mathrm{_{M}}$} &
\colhead{$M\mathrm{_{WD}}$} & \colhead{Age$\mathrm{_{WD}}$} & \colhead{$R\mathrm{_{WD}}$} &
\colhead{Sp$\mathrm{_{M}}$} & \colhead{$M\mathrm{_{M}}$} & \colhead{$R\mathrm{_{M}}$} \\
&(K)&(dex)&(K)&(dex)&(dex)&($M_{\Sun}$)&(Myr)&($R_{\Sun}$)&&($M_{\Sun}$)&($R_{\Sun}$)}
\startdata
J014141.83$+$311852.9 & 21000 & 8.25 & 3700 & 4.0 & 0.00 &  0.78 & 101.6 & 0.011 & 1.0 & 0.464 & 0.480 \\
J081934.63$+$024802.7 & 32000 & 7.00 & 3600 & 3.5 & $-$0.25 & 0.33 & 5.0 & 0.030 & 1.5 & 0.450 & 0.465 \\
J084006.40$+$143025.9 & 38000 & 8.50 & 3500 & 4.0 & $-$0.50 & 0.96 & 16.4 & 0.009 & 2.0 & 0.431 & 0.445 \\
J085900.86$+$493519.8 & 32000 & 8.50 & 3500 & 3.5 & $-$0.75 & 0.95 & 36.4 & 0.009 & 2.0 & 0.431 & 0.445 \\
J101616.82$+$310506.5 & 25000 & 8.25 & 3600 & 4.5 & $-$0.25 & 0.78 & 50.1 & 0.011 & 1.5 & 0.450 & 0.465 \\
J111035.16$+$280733.2 & 24000 & 7.75 & 3700 & 4.0 & $-$0.25  & 0.51 & 18.9 & 0.016 & 1.0 & 0.464 & 0.480 \\
J114732.25$+$593921.9 & 22000 & 8.25 & 3700 & 5.5 & 0.50 & 0.78 & 84.5 & 0.011 & 1.0 & 0.464 & 0.480 \\
J120024.56$+$292310.3 & 48000 & 7.75 & 3700 & 4.5 & $-$0.25 & 0.58 & 2.3 & 0.017 & 1.0 & 0.464 & 0.480 \\
J221102.56$-$002433.5 & 23000 & 8.50 & 3600 & 5.0 & 0.25 &  0.94 & 126.2 & 0.009 & 1.5 & 0.450 & 0.465 \\
J232004.01$+$270623.7 & 34000 & 7.50 & 3300 & 4.0 & 0.00 & 0.45 & 5.0 & 0.020 & 3.5 & 0.350 & 0.359
\enddata
\tablecomments{
Typical uncertainties of parameters have been mentioned in the text.}
\end{deluxetable}

\begin{deluxetable}{lrccccc}
\tablecolumns{7}
\tabletypesize{\scriptsize}
\setlength{\tabcolsep}{0.05in}
\tablewidth{0pt}
\tablecaption{Radial Velocities of LAMOST WDMS Binaries.}
\tablehead{
\colhead{Designation} & \colhead{HJD} &
\colhead{RV$\mathrm{_{H\alpha}}$} & \colhead{err$\mathrm{_{RV_{H\alpha}}}$} &
\colhead{RV$\mathrm{_{\Ion{Na}{I} D}}$} & \colhead{err$\mathrm{_{RV_{\Ion{Na}{I} D}}}$} &
\colhead{Flag} \\
&&(km s$^{-1}$)&(km s$^{-1}$)&(km s$^{-1}$)&(km s$^{-1}$)&}
\startdata
J014141.83$+$311852.9	&	2455918.0249		&	...		&	...		&	13.5		&	13.1		& a \\
J052531.26$+$283807.6	&	2455951.0953		&	60.7		&	12.6		&	18.6		&	13.7		& a \\
J073128.30$+$264353.8	&	2455921.2338		&	...		&	...		&	$-$102.4	&	15.0		& a \\
J081934.63$+$024802.7	&	2455921.3048		&	42.9		&	12.7		&	23.7		&	11.7		& a \\
						&	2455921.2920		&	49.1		&	11.9		&	23.3		&	11.1		& b \\
						&	2455921.3176		&	42.0		&	12.9		&	21.9		&	12.0		& b \\
J084006.40$+$143025.9	&	2455977.1160		&	48.7		&	14.1		&	64.9		&	14.5		& a \\
J085900.86$+$493519.8	&	2456021.0491		&	...		&	...		&	$-$4.8		&	11.7		& a \\
						&	2456021.0432		&	...		&	...		&	$-$5.7		&	10.4		& b \\
						&	2456021.0551		&	...		&	...		&	$-$0.9		&	10.4		& b \\
                   		&	2456021.0801		&	$-$33.9	&	16.5		&	4.02		&	10.6		& a \\
						&	2456021.0742		&	...		&	...		&	1.9		&	10.4		& b \\
						&	2456021.0859		&	...		&	...		&	3.8		&	10.5		& b \\
J101616.82$+$310506.5	&	2455907.3713		&	$-$16.6	&	12.7		&	$-$3.8		&	16.3		& a \\
						&	2455907.3624		&	$-$11.2	&	11.3		&	$-$0.3		&	11.0		& b \\
						&	2455907.3802		&	$-$11.6	&	12.7		&	$-$2.4		&	11.0		& b \\
                   		&	2455959.2835		&	$-$17.1	&	12.7		&	2.0		&	11.1		& a \\
                   		&	2455959.2681		&	$-$14.0	&	12.2		&	9.1		&	10.6		& b \\
                   		&	2455959.2835		&	$-$6.3		&	14.0		&	$-$8.7		&	12.0		& b \\
                   		&	2455960.3033		&	$-$19.8	&	11.6		&	$-$7.4		&	11.5		& a \\
                   		&	2455960.2868		&	$-$10.9	&	10.6		&	$-$1.2		&	10.4		& b \\
                   		&	2455960.3036		&	...		&	...		&	$-$21.3	&	11.6		& b \\
                   		&	2455960.3198		& 	...		&	...		&	3.0		&	11.4		& b \\
                   		&	2455978.1927		&	$-$13.3	&	13.7		&	$-$20.6	&	13.8		& a \\
                   		&	2455978.1676		& 	...		&	...		&	$-$11.5	&	11.7		& b \\
						&	2455978.1846		&	$-$28.4	&	14.8		&	4.2		&	12.0		& b \\
						&	2455978.2177		&	$-$8.5		&	13.1		&	$-$10.1	&	12.5		& b \\
J105405.25$+$283916.7	&	2455931.3041		&	10.3		&	12.1		&	...		&	...		& a \\
						&	2455931.2880		&	3.3		&	12.9		&	10.6		&	11.3		& b \\
						&	2455931.3031		&	25.9		&	11.9		&	11.5		&	11.0		& b \\
						&	2455931.3203		&	15.7		&	11.6		&	4.4		&	11.4		& b \\
						&	2453826.8103		&	30.0		&	15.0		&	32.4		&	10.4		& c \\
						&	2453826.7958		&	35.2		&	15.2		&	...		&	...		& c \\	
						&	2453826.8086		&	35.2		&	15.1		&	...		&	...		& c \\
						&	2453826.8231		&	33.3		&	15.2		&	...		&	...		& c \\
						&	2454887.2906		&	33.7		&	14.8		&	25.1		&	10.1		& c \\
J105421.88$+$512254.1	&	2455915.3897		&	33.0		&	15.4		&	23.0		&	12.9		& a \\
						&	2455915.4032		&	6.8		&	12.0		&	26.9		&	14.2		& b \\
						&	2452669.8635		&	31.9		&	14.5		&	31.5		&	10.8		& c \\
						&	2452669.8635		&	34.8		&	10.5		&	...		&	...		& c \\	
						&	2453759.6249		&	$-$60.8	&	4.2		&	...		&	...		& c \\
						&	2453759.7157		&	$-$42.9	&	4.3		&	...		&	...		& c \\
						&	2452345.3158		&	$-$54.6	&	14.7		&	$-$67.6	&	10.7		& c \\
J111035.16$+$280733.2	&	2455931.3041		&	...		&	...		&	$-$13.1	&	14.9		& a \\
J112512.36$+$290545.4	&	2456018.1430		&	$-$117.2	&	18.3		&	...		&	...		& a \\
						&	2453794.9024		&	$-$52.0	&	15.3		&	$-$41.1	&	10.8		& c \\
						&	2453794.8889		&	$-$64.8	&	17.4		&	...		&	...		& c \\	
						&	2453794.9047		&	$-$50.6	&	15.6		&	...		&	...		& c \\
						&	2453794.9182		&	$-$46.9	&	16.7		&	...		&	...		& c \\
J114732.25$+$593921.9	&	2455931.3750		&	...		&	...		&	$-$121.0	&	10.9		& a \\
						&	2455931.3485		&	...		&	...		&	$-$115.9	&	16.6		& b \\
						&	2455931.3739		&	...		&	...		&	$-$122.9	&	11.9		& b \\
						&	2455931.4016		&	...		&	...		&	$-$140.6	&	11.5		& b \\
J122037.01$+$492334.0	&	2455923.4110		&	...		&	...		&	$-$91.4	&	11.2		& a \\
						&	2452413.6647		&	$-$25.3	&	15.5		&	$-$23.8	&	11.2		& c \\
						&	2452413.6465		&	$-$11.4	&	19.1		&	...		&	...		& c \\
						&	2452413.6625		&	$-$33.3	&	17.0		&	...		&	...		& c \\
						&	2452413.6810		&	$-$37.7	&	17.2		&	...		&	...		& c \\
J135635.32$+$084841.7	&	2456062.1763		&	...		&	...		&	23.3		&	15.1		& a \\
J162617.40$+$384029.9	&	2455999.3847		&	$-$25.5	&	11.5		&	$-$16.4	&	10.9		& a \\
						&	2455999.3693		&	$-$22.0	&	13.4		&	$-$18.7	&	10.9		& b \\
						&	2455999.3850		&	$-$27.0	&	11.9		&	$-$18.5	&	11.5		& b \\
						&	2455999.4001		&	$-$32.4	&	11.7		&	$-$13.2	&	13.7		& b \\
J224609.42$+$312912.2	&	2455863.0592		&	$-$18.0	&	10.7		&	$-$3.6		&	10.7		& a \\
						&	2455863.0417		&	$-$18.3	&	10.8		&	$-$12.6	&	10.9		& b \\
						&	2455863.0582		&	$-$19.4	&	10.8		&	$-$4.0		&	11.6		& b \\
						&	2455863.0767		&	$-$8.3		&	11.1		&	$-$7.8		&	26.2		& b \\
J232004.01$+$270623.7	&	2455878.0210		&	$-$87.9	&	13.7		&	...		&	...		& a \\
\enddata
\tablecomments{
HJD is the heliocentric corrected date of the observations.  Flags a and b represent the radial velocities measured from the LAMOST combined spectra and LAMOST subspectra respectively. Flag c represents the radial velocities provided by \citet{rebassa2012a}.}
\end{deluxetable}

\begin{deluxetable}{lrcccccc}
\tablecolumns{5}
\tabletypesize{\scriptsize}
\setlength{\tabcolsep}{0.05in}
\tablewidth{0pt}
\tablecaption{Upper Limits to the Orbital Periods of the Two PCEB Candidates.}
\tablehead{
\colhead{Designation} &
\colhead{$P(\chi^2)$} &
\colhead{$M\mathrm{_{WD}}$} &
\colhead{$M\mathrm{_{M}}$} &
\colhead{$P\mathrm{_{orb}} < $} \\
&&($M_{\Sun}$)&($M_{\Sun}$)&(d)}
\startdata
J105421.88$+$512254.1	&	1.00000	&	0.67		&	0.38		&	2.71		\\
J122037.01$+$492334.0	&	0.99998	&	0.42		&	0.464	&	2.96		\\
\enddata
\end{deluxetable}

\end{document}